\documentclass[12pt]{article}

\usepackage{graphicx}

\input{epsf}

\def\rf#1{(\ref{eq:#1})}
\def\lab#1{\label{eq:#1}}

\def\br{\begin{eqnarray}}
\def\er{\end{eqnarray}}
\def\be{\begin{equation}}
\def\ee{\end{equation}}
\def\({\left(}
\def\){\right)}

\def\rlx{\relax\leavevmode}
\def\vp{\varphi}

\def\ve{\varepsilon}
\def\calb{{\cal B}}

\newcommand{\sbr}[2]{\left\lbrack\,{#1}\, ,\,{#2}\,\right\rbrack}

\def\IZ{\rlx\hbox{\sf Z\kern-.4em Z}}
\def\IR{\rlx\hbox{\rm I\kern-.18em R}}
\def\IC{\rlx\hbox{\,$\inbar\kern-.3em{\rm C}$}}
\def\one{\hbox{{1}\kern-.25em\hbox{l}}}

\begin{document}

\begin{titlepage}
\vspace*{-1cm}

\vskip 2cm

\vspace{.2in}
\begin{center}
{\large\bf A mild source for the Wu-Yang magnetic monopole}
\end{center}

\vspace{.5cm}

\begin{center}
C. P. Constantinidis~$^{\dagger}$, L. A. Ferreira~$^{\star}$ and G. Luchini~$^{\dagger}$

\vspace{.3 in}
\small

\par \vskip .2in \noindent 
$^{\dagger}$ Departamento de F\'isica\\
 Universidade Federal do Esp\'irito Santo (UFES),\\
 CEP 29075-900, Vit\'oria-ES, Brazil
 
\par \vskip .2in \noindent 
$^{\star}$ Instituto de F\'\i sica de S\~ao Carlos; IFSC/USP;\\
Universidade de S\~ao Paulo; USP  \\ 
Caixa Postal 369, CEP 13560-970, S\~ao Carlos-SP, Brazil\\

\vspace{2cm}

\normalsize
\end{center}


\begin{abstract}

\noindent We establish that the Wu-Yang monopole needs the introduction of  a magnetic point source at the origin in order for it to be a solution of  the differential and integral equations for the Yang-Mills theory. That result is corroborated by the analysis through distribution theory, of the two types of magnetic fields relevant for the local and global  properties of the Wu-Yang solution. The subtlety lies on the fact that with the non-vanishing magnetic point source required by the Yang-Mills integral equations, the Wu-Yang monopole configuration does not violate, in the sense of distribution theory, the differential Bianchi identity.

\end{abstract} 
\end{titlepage}

\section{Introduction}
\label{sec:intro}
\setcounter{equation}{0}

The purpose of this paper is to settle a long standing problem concerning the nature of the singularity of the magnetic monopole solution constructed in 1969 by T.T. Wu and C.N. Yang \cite{wuyang69} for the pure $SU(2)$  Yang-Mills differential field equations.  The solution presents a spherically symmetric non-abelian magnetic field with a  strength that depends on the inverse of the square of the radial distance, for all distance scales, and so it is singular at the origin. Such  singularity has been an issue since then, even though very few authors tried to concretely address the problem. In 1975 Wu and Yang comment, in the final section of their paper  \cite{wuyang75}, that their solution perhaps does not satisfy the Bianchi identity and so the corresponding field is not a proper gauge field at the origin. Lanyi and Pappas were perhaps the only authors to address that problem directly in their 1977 paper \cite{pappas}. Their results however, as they say, are inconclusive since they got different results using different methods. We show here that one of their results is not in fact correct. Some other authors   \cite{oh,butera} have discussed the necessity of sources for some types of singular solutions, but the Wu-Yang solution itself was not directly addressed.

In this paper we discuss how the integral Yang Mills equations,   proposed in  \cite{ym1,ym2}, shed a light on this question by revealing that the concerns presented in \cite{wuyang75} with respect to the compatibility of the Wu-Yang configuration with the Bianchi identity are indeed very strong and we show how those integral equations establish that the introduction of a point source (uniquely fixed by them) indeed solves this long standing question. The crucial and interesting aspect is that  with such a non-vanishing magnetic point source required by the Yang-Mills integral equations, the Wu-Yang monopole configuration does not violate, in the sense of distribution theory, the differential Bianchi identity, and that is why we call it a mild source.  

Another intriguing point concerning the Wu-Yang monopole solution is that so far it did not really possess a magnetic charge associated to it. Indeed, being a solution of the pure Yang-Mills theory without a Higgs field, it does not possess, like the 't Hooft-Polyakov monopole does \cite{thooft,polyakov}, a topological charge that can be interpreted as a magnetic charge. In addition, the usual (dynamically conserved) Noether magnetic charge of Yang-Mills theories vanishes when evaluated on the  Wu-Yang monopole solution \cite{shnir}. We show in this paper that the Wu-Yang monopole does possess a  dynamically conserved non-vanishing magnetic charge. It is constructed through the integral equations for the Yang-Mills theory  \cite{ym1,ym2}, and its conservation comes from an iso-spectral time evolution of some special operators, in a manner similar to what happens in integrable field theories, and so it is  not a Noether charge. In addition,  contrary to the usual  Noether magnetic charge, such charge is invariant under general (large) gauge transformations.  

In order to address the singularity issue of the Wu-Yang monopole we use the recently proposed integral equations for the Yang-Mills theory \cite{ym1,ym2} to fix in a unique way the type of point source one needs to introduce to make the solution consistent. The result obtained is then corroborated by the use of distribution theory in the differential Yang-Mills equations. The result we find is that the magnetic field of the Wu-Yang monopole solution must satisfy 
\be
{\vec D}\cdot  {\vec B}=-\frac{1}{e}\, \frac{{\hat r}\cdot {\vec T}}{r^2}\, \delta\(r\)
\lab{result}
\ee
where $\delta\(r\)/r^2$ is the radial part of the three dimensional Dirac delta function $\delta^{(3)}\({\vec r}\)$, $e$ is the gauge coupling constant, ${\hat r}={\vec r}/r$ is the unit vector in the radial direction, and ${\hat r}\cdot {\vec T}={\hat r}_a T_a$, with $T_a$ being the generators of the $SU(2)$ Lie algebra, i.e. $\sbr{T_a}{T_b}=i\,\ve_{abc}T_c$, $a,b,c=1,2,3$.  The magnetic field is defined as $B_i=-\frac{1}{2}\ve_{ijk}\,F_{jk}$, with the field tensor being $F_{\mu\nu}=\partial_{\mu} A_{\nu}-\partial_{\nu} A_{\mu}+i\,e\,\sbr{A_{\mu}}{A_{\nu}}$, $\mu ,\nu=0,1,2,3$, and the covariant derivative being $D_{\mu}\star=\partial_{\mu}\star+i\,e\,\sbr{A_{\mu}}{\star}$. With such a notation the Wu-Yang solution \cite{wuyang69} reads
\br
A_i= -\frac{1}{e}\,\ve_{ija}\,\frac{x^j}{r^2}\,T_a\;, \qquad\qquad\qquad
F_{ij}=\frac{1}{e}\,\ve_{ijk}\,\frac{x^k}{r^3}\,{\hat r}\cdot {\vec T}
\lab{wuyangsol}
\er
with $A_0=0$ and $F_{0i}=0$. 
The fact that the point source in \rf{result} contains only the radial part of the Dirac delta function  will prove to be crucial for the compatibility between the results about the analyticity of the solution, obtained through the integral equations of Yang-Mills theory and that of distribution theory applied to the differential Yang-Mills equations. In addition, it makes the source spherically symmetric  under the joint action of physical space and isospin space rotations.

The analysis of the singularity of the Wu-Yang monopole solution has to take into account the fact that there are two types of magnetic fields relevant for its physical properties. The first one is the usual magnetic field of Yang-Mills theory,  the Hodge dual of the space components of the field tensor, i.e. $B_i=-\frac{1}{2}\ve_{ijk}\,F_{jk}$. The second magnetic field appears in the context of the integral equations for the Yang-Mills theory \cite{ym1,ym2}.  In those equations the Yang-Mills field tensor, as well as its Hodge dual, always appear conjugated by the Wilson line operator $W$ i.e.  $F_{\mu\nu}^W\equiv W^{-1}\, F_{\mu\nu}\,W$, where $W$ is integrated along a path, starting at a given reference point and ending at the point where $F_{\mu\nu}$ is evaluated. Those paths are defined by the scanning of surfaces and volumes used in the integral equations. Therefore, the second type of magnetic field that we have is   the usual non-abelian magnetic field $B_i$ (the Hodge dual of $F_{ij}$), conjugated by the Wilson line operator, i.e.  $B_i^W\equiv W^{-1}\, B_i\,W$. In the case of the Wu-Yang monopole the conjugation by the Wilson operator renders the second magnetic field $B_i^W$ lying in an abelian $U(1)$ subalgebra of the $SU(2)$ Lie algebra, and its properties differ drastically from those of $B_i$, and that plays a crucial role in our analysis. We now discuss these two types of magnetic fields.  

\section{The first type of magnetic field}
\label{sec:firstmagfield}
\setcounter{equation}{0}

The first type of magnetic field that we consider is the usual Hodge dual of the spatial part of the field tensor, i.e.  $B_i=-\frac{1}{2}\ve_{ijk}\,F_{jk}$, and so from the Wu-Yang monopole solution \rf{wuyangsol} we have 
\be
{\vec B}=-\frac{1}{e}\,\frac{{\hat r}}{r^2} \;{\hat r}\cdot {\vec T}
\lab{wuyangmagfield}
\ee
Note that in fact, we have here  three magnetic fields, since  expanding it on a basis $T_a$, $a=1,2,3$, of the $SU(2)$ Lie algebra of the gauge group,  one gets ${\vec B}={\vec B}_a\,T_a$, with ${\vec B}_a= -\frac{1}{e}\frac{{\hat r}}{r^2} \;{\hat r}_a$. 
We have  dynamically conserved magnetic charges associated to such  fields, that are obtained from the Noether's theorem applied to the gauge symmetry of the Yang-Mills action. However, it is more convenient to get them directly from the differential Yang-Mills equations 
\be
D_{\nu} F^{\nu\mu}= J^{\mu}\;\qquad\qquad\qquad\qquad 
D_{\nu}{\widetilde F}^{\nu\mu}= j^{\mu}
\lab{diffymeqs}
\ee
where ${\widetilde F}_{\mu\nu}$ is the Hodge dual of the field tensor, i.e. ${\widetilde F}_{\mu\nu}\equiv \frac{1}{2}\,\varepsilon_{\mu\nu\rho\lambda}\, F^{\rho\lambda}$,  and 
where we allowed for a magnetic current $j_{\mu}$, besides the usual matter current $J_{\mu}$. Indeed, writing the second set of Yang-Mills equations \rf{diffymeqs} as
\be
{\tilde K}^{\mu}\equiv \partial_{\nu}{\tilde F}^{\nu\mu}=-i\,e\,\sbr{A_{\nu}}{{\tilde F}^{\nu\mu}}+j^{\mu}
\lab{magcurrdef}
\ee
one gets that, due to the antisymmetry of the dual of the field tensor, the current ${\tilde K}^{\mu}$,  has a vanishing divergence everywhere, i.e. $\partial_{\mu}{\tilde K}^{\mu}=0$,  except perhaps at the origin where the ordinary derivatives might not commute due to the singularity. We have that 
\be
{\tilde K}_0=-\partial_iB_i=j_0\;;\qquad\qquad\qquad\qquad  {\tilde K}_i=-\partial_0B_i=j_i=0
\lab{j0divegencerel}
\ee
 The fact that the spatial part of the magnetic current, ${\vec j}$, has to vanish follows from the Yang-Mills equations themselves, since $j^i=D_j {\widetilde F}^{ji} + D_0{\widetilde F}^{0i}=0$, because ${\widetilde F}^{ji}=0$ (no electric field), and  $D_0{\widetilde F}^{0i}=0$, since $A_0=0$, and the solution is static. Note that for the solution \rf{wuyangsol} the commutator term in \rf {magcurrdef}  vanishes, since $A_0=0$ and ${\tilde F}_{ij}=0$, and so $\sbr{A_{\nu}}{{\tilde F}^{\nu i}}=0$. In addition, one has $\sbr{A_{\nu}}{{\tilde F}^{\nu 0}}=-\sbr{A_i}{B_i}=-\frac{1}{e^2}\,\ve_{ija}\,\frac{{\hat r}^j}{r}\,\frac{{\hat r}^i}{r^2}\,\sbr{T_a}{{\hat r}\cdot {\vec T}}=0$, since $\ve_{ija} \,{\hat r}^i\,{\hat r}^j=0$. Note also that the gauge field $A_i$ has a singularity at the origin of the form $1/r$ and the magnetic field $B_i$ a singularity of the form $1/r^2$. Therefore, they are both locally summable functions in three space dimensions (see \cite{distribution} for details on that).  Their product  produces terms with singularities of the form $1/r^3$, and therefore these are not locally summable in three dimensions. However, the sum in the spatial index $i$ in the commutator 
$\sbr{A_i}{B_i}$, cancels out those terms, i.e. we have that
\be
\sbr{A_i}{B_i}=0\;;\qquad\qquad\qquad \mbox{\rm and so}\qquad \qquad\qquad D_iB_i=\partial_iB_i
\lab{nullcommutatorab}
\ee 
Therefore, from the point of view of distribution theory such a  commutator term  does not present problems for our analysis.

\begin{figure*}[h]
\begin{center}
		\includegraphics[width=7cm]{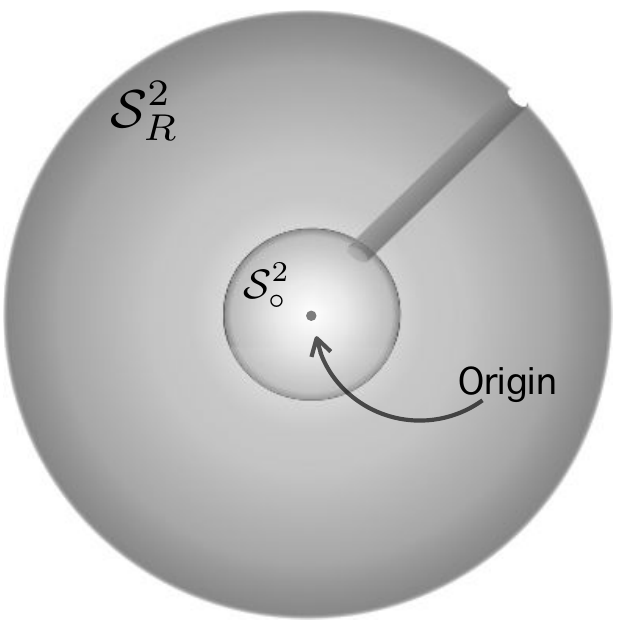}
		\caption{Surface $\partial\Omega$ made of two spheres centered at the origin and joined by a thin radial cylinder.
			 }
		\label{fig:surfaceV}
\end{center} 
	\end{figure*}

Thus, considering  a three-volume $\Omega$ that does not contain the origin, we have that the cor\-res\-pon\-ding conserved magnetic charges are given by
\be
{\tilde Q}_\Omega=\int_{\Omega} d^3x\, {\tilde K}^{0} =\int_{\Omega} d^3x\,j_0 = -\int_{\Omega} d^3x\, \partial_iB_i=-\int_{\partial \Omega}d\Sigma_i\,B_i
\lab{magchargedef}
\ee
where in the last equality we have used Gauss theorem (abelian Stokes theorem) to each one of the three Lie algebra components $B_i^a$ of the magnetic field $B_i=B_i^a\,T_a$. Let us now consider a volume $\Omega$ as shown in Figure \ref{fig:surfaceV} with a border $\partial \Omega$ made of three surfaces, namely, two spheres centered at the origin, one of them $S_R^2$ with a very large radius, and the second one $S_0^2$ with a very small radius, and a very thin radial cylinder joining them.  Therefore, the origin is not inside the volume $\Omega$.  Using spherical polar coordinates, where $x^1=r\,\sin\theta\,\cos\vp$, $x^2=r\,\sin\theta\,\sin\vp$, and $x^3=r\,\cos\theta$,  we have that
\be
{\hat r}\cdot {\vec T}=\sin\theta\(\cos\vp\,T_1+\sin\vp\,T_2\)+\cos\theta\,T_3
\lab{rtpolar}
\ee
and so 
\be
\int_{S_R^2}d\Sigma_i\,B_i= -\frac{1}{e}\,\int_0^{\pi} d\theta\,\int_0^{2\pi}d\vp\, \sin\theta\,\, {\hat r}\cdot {\vec T}=0
\lab{vanishreasoning}
\ee 
For the same reason we have that the surface integral on $S_0^2$ (oriented inwards) vanishes, i.e. $\int_{S_0^2}d\Sigma_i\,B_i=0$. The surface integral on the radial cylinder vanishes because the magnetic field $B_i$ is radial and the area element $d\Sigma_i$ on it, is perpendicular to the radial direction. That can be repeated on any volume $\Omega$ not containing the origin, and the magnetic flux on its border will always vanish. Therefore, from \rf{magchargedef} one sees that the source $j_0$ has to vanish everywhere except perhaps at the origin. 

Given the singularity of the magnetic field \rf{wuyangmagfield} at the origin, we now evaluate  its   ordinary divergence  using distribution theory \cite{distribution}. The magnetic field \rf{wuyangmagfield} has a singularity $1/r^2$ at the origin and so its components $B_i$ are locally summable functions in $\IR^3$. Therefore, we can define the distributions  
\be
\langle T_{B_i},\Phi\rangle \equiv \int d^3 x\, B_i\, \Phi
\lab{distributionbi}
\ee
where the test functions $\Phi$ are $C^{\infty}$ functions that vanish outside a given compact region around the origin. The distributions \rf{distributionbi} are linear functionals on the vector space defined by the test functions $\Phi$. The derivatives of such  distribution are defined as \cite{distribution}
\be
\langle \frac{\partial T_{B_i}}{\partial x_j},\Phi\rangle \equiv -\langle  T_{B_i},\frac{\partial\Phi}{\partial x_j}\rangle=-\int d^3 x\, B_i\, \frac{\partial\Phi}{\partial x_j} 
\lab{derivativedistrib}
\ee
Even though the functions $B_i$ may not have well defined derivatives at the origin, the distributions $\langle T_{B_i},\Phi\rangle$ do have, since the partial derivatives in \rf{derivativedistrib} act on the  $C^{\infty}$ test functions $\Phi$. In addition, different derivatives commute when acting on $\langle T_{B_i},\Phi\rangle$.  The distribution associated to the divergence of the magnetic field \rf{wuyangmagfield} is then obtained from \rf{derivativedistrib} by taking $i=j$, and summing over $i$, i.e. 
\be
\langle T_{{\vec \nabla}\cdot {\vec B}},\Phi\rangle =\int d^3{\vec r}\, {\vec \nabla}\cdot {\vec B}\; \Phi = - \int d^3{\vec r}\,  {\vec B}\cdot {\vec \nabla}\Phi
\lab{graddef}
\ee
 Note that the test functions $\Phi$ are not Lie algebra valued, but are ordinary scalar test functions. The reason is that by expanding  the magnetic field on a basis $T_a$, $a=1,2,3$, for the $SU(2)$ Lie algebra as ${\vec B}= {\vec B}_a\,T_a$, one observes that \rf{graddef} corresponds in fact to the three equations 
$ \int d^3{\vec r}\, {\vec \nabla}\cdot {\vec B}_a\; \Phi = - \int d^3{\vec r}\,  {\vec B}_a\cdot {\vec \nabla}\Phi$, for $a=1,2,3$. 

Again using spherical polar coordinates  we have that only the radial part of ${\vec \nabla}\Phi$ contributes to the r.h.s. of \rf{graddef}, and so 
 \br
\int d^3{\vec r}\, {\vec \nabla}\cdot {\vec B}\; \Phi   
 =\frac{1}{e}\,\int d\theta\,d\vp\; \sin\theta\;{\hat r}\cdot {\vec T}\;\left[\Phi\(\infty\,,\theta\,,\,\vp\)-\Phi\(0\,,\theta\,,\,\vp\)\right].
\er
But $\Phi\(\infty\,,\theta\,,\,\vp\)=0$, since $\Phi$ vanishes outside the compact region, and $\Phi\(0\,,\theta\,,\,\vp\)=\Phi\({\vec 0}\)$. Therefore, using the same reasoning as in \rf{rtpolar} and \rf{vanishreasoning} one gets 
\be
\int d^3{\vec r}\, {\vec \nabla}\cdot {\vec B}\; \Phi= -\frac{1}{e}\,\Phi\({\vec 0}\)\,\int d\theta\,d\vp\; \sin\theta\;{\hat r}\cdot {\vec T}=0.
\lab{divb=0}
\ee
According to  distribution theory one has to find a locally summable function $u$ such that
\be
\int d^3{\vec r}\, u\; \Phi=\int d^3{\vec r}\, {\vec \nabla}\cdot {\vec B}\; \Phi =0
\lab{udef}
\ee
which leads to the result ${\vec \nabla}\cdot {\vec B}=u$. Obviously the trivial function $u=0$ does the job. However, it is not the only one. Consider locally summable functions of the type 
\be
u= C\, \frac{{\hat r}\cdot {\vec T}}{r^2}\,\delta\(r\)
\lab{generalu}
\ee
 where $C$ is a constant. Then one gets 
\br
\int d^3{\vec r}\, u\; \Phi=C \int dr\,d\theta\,d\vp\; r^2\,\sin\theta\;
 \frac{{\hat r}\cdot {\vec T}}{r^2}\;\delta\(r\)\,\Phi
 =C\,\Phi\({\vec 0}\)\,\int d\theta\,d\vp\; \sin\theta\;
{\hat r}\cdot {\vec T}=0
\lab{checkgeneralu}
\er
where in the last equality we have used the same reasoning as in  \rf{divb=0}. 

According to distribution theory \cite{distribution} if the integral of a locally summable function $u$ multiplied by any test function $\Phi$ gives zero, then $\langle T_u,\Phi\rangle$ is the zero distribution. Therefore, the distributions associate to any of the functions \rf{generalu}  are the same as the zero distribution. If two locally summable functions $u_1$ and $u_2$ lead to the same distributions, i.e. $\langle T_{u_1},\Phi\rangle=\langle T_{u_2},\Phi\rangle$, then $u_1$ and $u_2$ can differ on a set of zero measure only \cite{distribution}, and that is what happens with the functions \rf{generalu}.  Consequently, from this point of view, the distribution associated to the divergence of the magnetic field \rf{wuyangmagfield} is the zero distribution. Therefore, the divergence of that magnetic field, and so the source $j_0$, can differ from zero only on a set of  zero measure. The fact that they do not vanish in a set of zero measure will prove important in the global analysis of the Wu-Yang solution in the next section. 

One can then extend the definition of the conserved magnetic charges, introduced in \rf{magchargedef}, to any volume containing the origin, and they will vanish. Indeed, one has that
\br
{\tilde Q}_{\IR^3}&=&\int_{\IR^3} d^3x\, {\tilde K}^{0} =\int_{\IR^3} d^3x\,j_0 = -\int_{\IR^3} d^3x\, \partial_iB_i
=-\int_{\IR^3} d^3x\, C\, \frac{{\hat r}\cdot {\vec T}}{r^2}\,\delta\(r\)=0
\nonumber\\
&=&-\int_{S^2_{\infty}}d\Sigma_i\,B_i=
\frac{1}{e}\,\int_0^{\pi} d\theta\,\int_0^{2\pi}d\vp\, \sin\theta\,\, {\hat r}\cdot {\vec T}=0
\lab{magchargeallspace}
\er
where we have used the same reasoning as in \rf{rtpolar} and  \rf{vanishreasoning} in the  volume and surface integrals. 
Therefore, the Wu-Yang monopole does not have a non-trivial, dynamically conserved, magnetic charge associated to the magnetic field \rf{wuyangmagfield}. Note in addition, that the magnetic charges \rf{magchargedef} (or \rf{magchargeallspace}) are not gauge invariant, since under a gauge transformation $A_{\mu}\rightarrow g\,A_{\mu}\,g^{-1}  +\(i/e\) \partial_{\mu}g\,g^{-1}$, one has that ${\tilde Q}_{\IR^3} \rightarrow -\int_{S^2_{\infty}}d\Sigma_i\,g\,B_i\,g^{-1}$. For gauge transformations where $g$ goes to a constant element $g_0$ at spatial infinity, then ${\tilde Q}_{\IR^3} \rightarrow g_0\,{\tilde Q}_{\IR^3} \,g_0^{-1}$, and so the eigenvalues of ${\tilde Q}_{\IR^3}$ would be gauge invariant. But that is far from being true for general gauge transformations. The fact that such magnetic charge of the Wu-Yang monopole vanishes was already pointed out by Wu and Yang in their original paper \cite{wuyang69} (see also \cite{shnir}).   Before we turn our analysis to the second magnetic field relevant for the physics of the Wu-Yang magnetic monopole, let us discuss a further argument supporting the results above. 

\subsection{The analysis by the smearing of the fields}

Following \cite{pappas} we consider a smeared version of the fields of the Wu-Yang magnetic monopole \rf{wuyangsol} and \rf{wuyangmagfield} as 
\br
A_i^{\rm reg.}= -\frac{1}{e}\,g\(r,a\)\,\ve_{ija}\,{\hat r}^j\,T_a\;, \qquad\qquad\qquad
B_{i}^{\rm reg.}=-\frac{1}{e}\,G\(r,a\)\,{\hat r}_i\,{\hat r}\cdot {\vec T}
\lab{wuyangsolsmear}
\er
where $a$ is a  regularizing positive parameter, and the smearing functions $g$ and $G$ have to satisfy
\be
\displaystyle \lim_{a \to 0}\, g\(r,a\)=\frac{1}{r}\;;\qquad\qquad\qquad 
\displaystyle \lim_{r \to \infty}\,r\,g\(r,a\)=1
\lab{smeargdef}
\ee
and 
\be
\displaystyle \lim_{a \to 0}\,G\(r,a\)=\frac{1}{r^2}\;;\qquad\qquad\qquad 
\displaystyle \lim_{r \to \infty}\,r^2\,G\(r,a\)=1
\lab{smearggdef}
\ee
The same reasoning leading to \rf{nullcommutatorab}, also leads to $\sbr{A_i^{\rm reg.}}{B_{i}^{\rm reg.}}=0$, and so
\be
{\vec D}\cdot {\vec B}^{\rm reg.}={\vec \nabla}\cdot {\vec B}^{\rm reg.}=-\frac{1}{e}\,{\hat r}\cdot {\vec T}\,\(G^{\prime}+\frac{2\,G}{r}\)=-\frac{1}{e}\,{\hat r}\cdot {\vec T}\,\frac{H^{\prime}}{r^2}
\lab{regdivergent}
\ee
where primes denote derivatives w.r.t. $r$, and where we have introduced the function $H\(r,a\)\equiv r^2\,G\(r,a\)$.
Note that the choice
\be
G\(r,a\)=\frac{1}{r^2+a^2}\;;\qquad\qquad \mbox{\rm and so}\qquad\quad H\(r,a\)=\frac{r^2}{r^2+a^2}
\lab{particularsmear}
\ee
satisfy the conditions \rf{smearggdef}. We then have 
\be
H^{\prime}=\frac{2}{a}\,\frac{r/a}{\(1+\(r/a\)^2\)^2}
\ee
and so $H^{\prime}\(0,a\)=0$, $H^{\prime}\(r,a\)\rightarrow 0$, as $r\rightarrow \infty$, and $H^{\prime}\(r,a\)$ has a maximum at $r=a/\sqrt{3}$. Consequently, as $a\rightarrow 0$, we have that $H^{\prime}$ diverges and its peak moves to $r=0$. In addition, we have that
\be
\int_0^{\infty}dr\, H^{\prime}=1\;;\qquad\qquad\qquad \mbox{\rm for any value of $a$}
\ee
Therefore, we conclude that 
\be
\displaystyle \lim_{a \to 0}\,H^{\prime}\(r,a\)= \delta\(r\)\;\qquad\qquad{\rm with}\qquad \qquad \int_0^{\infty}dr\,\delta\(r\) =1
\ee
and where $\delta\(r\)/r^2$ is the radial part of the Dirac delta function. So, one then gets from \rf{regdivergent} that
\be
\displaystyle \lim_{a \to 0}\,{\vec D}\cdot {\vec B}^{\rm reg.}=-\frac{1}{e}\,{\hat r}\cdot {\vec T}\,\frac{\delta\(r\)}{r^2}
\lab{smearresult}
\ee
There are three main conclusions to take from such a result. First, \rf{smearresult} is compatible with the analysis of section \ref{sec:firstmagfield}, using distribution theory, and leading to \rf{generalu}. Second, it fixes the value of the constant $C$, in \rf{generalu}, to $-1/e$, and so it makes \rf{smearresult} compatible with the main claim of this paper stated in \rf{result}. Third, it shows that the results of \cite{pappas}  are not quite correct. Indeed, in equation (11) of \cite{pappas} it is claimed that the source on the r.h.s. of \rf{smearresult} should be instead $- 4\,\pi\,{\hat r}\cdot {\vec T}\,\delta^{(3)}\({\vec r}\)$, where $\delta^{(3)}$ is the three-dimensional Dirac delta function. However, from \rf{regdivergent} one observes that the function multiplying ${\hat r}\cdot {\vec T}$, on its r.h.s., is a function of the radial distance only, and so one should not expect, in the limit $a\rightarrow 0$, a function depending upon the angles too. 

Of course, if one uses the regularized gauge potential $A_i^{\rm reg.}$, given in \rf{wuyangsolsmear}, to calculate the field tensor and so the magnetic field, one finds that the covariant divergence of that magnetic field vanishes,  since one is working with regular fields which certainly satisfy identities, namely the Bianchi identity. That is in fact the result expressed in equation (7b) of reference \cite{pappas}. Note that the regularized magnetic field $B_{i}^{\rm reg.}$, given in \rf{wuyangsolsmear}, is not that magnetic field obtained from the field tensor associated to the regularized gauge field $A_{i}^{\rm reg.}$, also given in \rf{wuyangsolsmear}, and those are the reasons behind the  ambiguous conclusions obtained in \cite{pappas}.

The result \rf{smearresult} was obtained from a particular choice of the regularizing function $G\(r,a\)$, given in \rf{particularsmear}, and a more general analysis is desirable. However, \rf{smearresult} alone already strengthens  the fact that \rf{result} is indeed correct. We now turn to a more general and more po\-wer\-ful global analysis  of the singularity of the Wu-Yang monopole using the integral equations of Yang-Mills theory \cite{afg1,afg2}.

\section{The second type of magnetic field}
\label{sec:secondmagfield}
\setcounter{equation}{0}

The second type of magnetic field appears when applying the integral Yang-Mills equations for the field configuration of the Wu-Yang magnetic monopole. Such  integral equations for the Yang-Mills theory were obtained in \cite{ym1,ym2} using a ge\-ne\-ra\-li\-za\-tion \cite{afg1,afg2} of the non-abelian Stokes theorem \cite{stokes} for a pair $\(\calb_{\mu\nu}\,,\,A_{\mu}\)$, of an antisymmetric tensor $\calb_{\mu\nu}$ and a one-form connection $A_{\mu}$, as follows. A more detailed discussion of such a theorem and the integral Yang-Mills equations can be found in the Appendix \ref{app:stokes}. Here we summarize the main results. In a space-time $M$ consider a three-volume $\Omega$ with border $\partial \Omega$. Choose a reference point $x_R$ on $\partial \Omega$ and scan $\Omega$ with closed surfaces, based on $x_R$, labelled by $\zeta$, such that $\zeta=0$ corresponds to the infinitesimal surface around $x_R$, and $\zeta=\zeta_0$ corresponds to $\partial \Omega$. Each closed surface scanning $\Omega$, is scanned in its turn by closed loops, based on $x_R$, labelled by $\tau$, such that $\tau=0$ and $\tau=2\,\pi$ correspond to the infinitesimal loops around $x_R$,  at the beginning and ending of the scanning. Each loop on its turn is parameterized by $\sigma$, starting and ending at $x_R$, such that $\sigma=0$ and $\sigma =2\,\pi$ correspond to the end points of the loop. The generalized non-abelian Stokes theorem \cite{afg1,afg2,ym1,ym2} states that
\be
V\(\partial\Omega\)\equiv P_2 \, e^{\int_{\partial\Omega}d\tau d\sigma \,W^{-1}\,\calb_{\mu\nu}\,W\, \frac{dx^{\mu}}{d\sigma}\,\frac{d\,x^{\nu}}{d\,\tau}}= P_3\, e^{\int_{\Omega} d\zeta \,{\cal K}}\equiv U\(\Omega\)
\lab{stokes}
\ee
where $P_3$ and $P_2$ mean volume and surface ordering respectively, according to the scanning described above. The quantity $V$, on the left, is obtained by integrating the equation
\begin{eqnarray}
\frac{d\,V}{d\,\tau}-V\, T\(\calb,A,\tau\)=0 \qquad \quad{\rm with}
\qquad\quad
T\(\calb,A,\tau\)\equiv
\int_{0}^{2\,\pi}d\sigma\; W^{-1}\,\calb_{\mu\nu}\,W\, \frac{d\,x^{\mu}}{d\,\sigma}\,\frac{d\,x^{\nu}}{d\,\tau} 
\lab{eqforvb}
\end{eqnarray} 
 where the $\sigma$-integration is along the loop  labeled by $\tau$, and $W$ is the Wilson line obtained by integrating, along the loop, the equation  
\be
\frac{d\,W}{d\,\sigma}+   i\,e\,A_{\mu}\,\frac{d\,x^{\mu}}{d\,\sigma}\,W=0.
\lab{eqforw}
\ee
On the other hand the quantity on the right, defined here as $U\(\Omega\)$, is obtained by integrating the equation 
\be
\frac{d\,U}{d\,\zeta} - {\cal K}\, U=0
\lab{eqforv2}
\ee
where ${\cal K}$ is given by 
\br
{\cal K}&\equiv&
\int_0^{2\,\pi}d\tau \int_0^{2\,\pi}d\sigma V\,\left\{ 
 W^{-1}\,
\left[D_{\lambda}\calb_{\mu\nu}+D_{\mu}\calb_{\nu\lambda}+D_{\nu}\calb_{\lambda\mu}\right]
\,W\frac{d\,x^{\mu}}{d\,\sigma}\,\frac{d\,x^{\nu}}{d\,\tau}\,
\frac{d\,x^{\lambda}}{d\,\zeta}\right.\nonumber\\
&-&\left. \int_0^{\sigma}d\sigma^{\prime}
\sbr{\calb_{\kappa\rho}^W\(\sigma^{\prime}\)-ie F_{\kappa\rho}^W\(\sigma^{\prime}\)}
{\calb_{\mu\nu}^W\(\sigma\)}\frac{dx^{\kappa}}{d\sigma^{\prime}}\frac{dx^{\mu}}{d\sigma}\right.
\nonumber\\
&&\left.\times
\(\frac{d\,x^{\rho}\(\sigma^{\prime}\)}{d\,\tau}\frac{d\,x^{\nu}\(\sigma\)}{d\,\zeta}
-\frac{d\,x^{\rho}\(\sigma^{\prime}\)}{d\,\zeta}\,\frac{d\,x^{\nu}\(\sigma\)}{d\,\tau}\)\right\} V^{-1}
\lab{calkdef}
\er  
where we have introduced the notation $X^W\equiv W^{-1}\,X\,W$. The integral equations for the Yang-Mills theory \cite{ym1,ym2} is obtained from such generalized non-abelian Stokes theorem by taking the connection $A_{\mu}$ to be the Yang-Mills gauge field  satisfying the differential Yang-Mills equations \rf{diffymeqs}, and the antisymmetric tensor $\calb_{\mu\nu}$ to be a linear combination of the field tensor and its Hodge dual, i.e.
\be
\calb_{\mu\nu}=i\,e\,\left[\alpha\,F_{\mu\nu}+\beta\,{\widetilde F}_{\mu\nu}\right]
\lab{calbdef}
\ee
The parameters $\alpha$ and $\beta$ are arbitrary and can in fact be even complex. One can expand both sides of the Yang-Mills integral equations \rf{stokes}, in powers of $\alpha$ and $\beta$, and obtain an infinite number of integral equations which must be satisfied by any solution. See \cite{directtest} for a detailed study in the case of the 't Hooft-Polyakov monopole. The quantity ${\cal K}$ given in \rf{calkdef} becomes
\br
{\cal K}&\equiv&
\int_0^{2\,\pi}d\tau \int_0^{2\,\pi}d\sigma V\,\left\{ 
 i\,e\,W^{-1}\,
\left[\alpha\,{\tilde j}_{\mu\nu\lambda}+ \beta\, {\tilde J}_{\mu\nu\lambda}\right]
\,W\frac{d\,x^{\mu}}{d\,\sigma}\,\frac{d\,x^{\nu}}{d\,\tau}\,
\frac{d\,x^{\lambda}}{d\,\zeta}\right.\nonumber\\
&+&\left. e^2\,\int_0^{\sigma}d\sigma^{\prime}
\sbr{\(\(\alpha-1\)\, F_{\kappa\rho}^W+\beta\, {\tilde F}_{\kappa\rho}^W\)\(\sigma^{\prime}\)}
{\(\alpha\, F_{\mu\nu}^W+\beta\, {\tilde F}_{\mu\nu}^W\)\(\sigma\)}\frac{dx^{\kappa}}{d\sigma^{\prime}}\frac{dx^{\mu}}{d\sigma}\right.
\nonumber\\
&&\left.\times
\(\frac{d\,x^{\rho}\(\sigma^{\prime}\)}{d\,\tau}\frac{d\,x^{\nu}\(\sigma\)}{d\,\zeta}
-\frac{d\,x^{\rho}\(\sigma^{\prime}\)}{d\,\zeta}\,\frac{d\,x^{\nu}\(\sigma\)}{d\,\tau}\)\right\} V^{-1}
\lab{calkym} 
\er 
where ${\tilde j}_{\mu\nu\lambda}$ and ${\tilde J}_{\mu\nu\lambda}$ are the Hodge duals of the magnetic and electric currents respectively, i.e.  $j^{\mu}=\frac{1}{3!}\ve^{\mu\nu\rho\lambda}\,{\tilde j}_{\nu\rho\lambda}$ and $J^{\mu}=\frac{1}{3!}\ve^{\mu\nu\rho\lambda}\,{\tilde J}_{\nu\rho\lambda}$.

As explained in \cite{ym1,ym2} the integral equations for the Yang-Mills theory lead in a quite natural way to the construction of dynamically conserved electric and magnetic charges. Such charges are obtained as eigenvalues of the  operator
\be
Q_{{\cal S}}=P_2 e^{ie\int_{\partial {\cal S}}d\tau d\sigma\, W^{-1}\, (\alpha F_{ij}+\beta {\widetilde F}_{ij}) \,W\,\frac{dx^{i}}{d\sigma}\frac{dx^{j}}{d\tau}}=  P_3e^{\int_{{\cal S}} d\zeta \;{\cal K}}\;,
\lab{chargeym}
\ee
where ${\cal S}$ is the spatial sub-manifold of the space-time under consideration. It is shown in \cite{ym1,ym2} that, under appropriated boundary conditions where the field tensor and the currents (magnetic and electric)  have to fall to zero as $1/r^{3/2+\delta^{\prime}}$ and $1/r^{2+\delta}$, respectively, with $\delta\;,\;\delta^{\prime}>0$, such an operator has an iso-spectral evolution in time, and so it eigenvalues are conserved. Those eigenvalues are the dynamically conserved charges of  the Yang-Mills theory, and contrary to the Noether charges  (see \rf{magchargedef}), they  are invariant under general gauge transformations.   As we show below the magnetic charge associated to \rf{chargeym} does not vanish when evaluated on the Wu-Yang monopole.

We now apply the integral Yang-Mills equations \rf{stokes} to the fields of the Wu-Yang magnetic monopole. In order to avoid problems with the singularity of that configuration we shall work with three-volumes $\Omega$ that do not contain the origin. So, we will be dealing with volumes of the type shown in Figure \ref{fig:surfaceV}.

The crucial property of the Wu-Yang solution, for our calculations, is that the field tensor conjugated by  the Wilson line, namely $W^{-1}\,F_{ij}\,W$, has a fixed direction in the Lie algebra, and so effectively becomes abelian.  That is a consequence of the fact that the Lie algebra element 
${\hat r}\cdot {\vec T}$ is covariantly constant. Indeed, one can check that for the connection \rf{wuyangsol} one has 
\be
i\,e\,\sbr{A_i}{{\hat r}\cdot {\vec T}}=\frac{1}{r}\left({\hat r}_i\, {\hat r}\cdot {\vec T}- T_i\right)
=-\partial_i \,{\hat r}\cdot {\vec T}
\lab{zerocovder}
\ee
and so $D_i \,{\hat r}\cdot {\vec T}=0$. Therefore, using \rf{eqforw} one gets 
\be
\frac{d\;}{d\sigma} \(W^{-1}\,{\hat r}\cdot {\vec T}\,W\)=W^{-1}\, \(D_i\, {\hat r}\cdot {\vec T}\)\, W\, \frac{d\,x^i}{d\,\sigma}=0.
\lab{rtconstant}
\ee
Consequently, the quantity $W^{-1}\,{\hat r}\cdot {\vec T}\,W$ is the same  in any point of a given loop, scanning the surfaces and volumes, and so equal to its value at the reference point $x_R$, where all the loops start and end. By fixing the integration constant in  \rf{eqforw} to be unity one gets that the Wilson line $W$ is unity at the reference point. Therefore 
\be
W^{-1}\,{\hat r}\cdot {\vec T}\,W= T_R;\qquad\qquad \mbox {\rm and so}\qquad \qquad
W^{-1}\,F_{ij}\,W=\frac{1}{e}\,\ve_{ijk}\,\frac{x^k}{r^3}\, T_R
\lab{fconjugated}
\ee
where $T_R \equiv {\hat r}_R\cdot {\vec T}$, and ${\hat r}_R$ is the unit radial vector at the reference point $x_R$, and where we have used  \rf{wuyangsol}. Note that  the derivatives in \rf{zerocovder} and \rf{rtconstant} do not involve the radial variable $r$, since the unit vector ${\hat r}$ does not depend upon it, and so we do not have problems with the singularity at the origin in the derivation of \rf{zerocovder} and \rf{rtconstant}. In addition, the derivatives of the Wilson line $W$ do not present problems either. Indeed, the evaluation of the Wilson operator even for loops that pass through the Wu-Yang monopole singularity presents no problem as explained in the appendix of \cite{ym1}. 

From \rf{fconjugated} we then come to the definition of the second magnetic field, as the conjugation, by the Wilson line, of the usual magnetic field \rf{wuyangmagfield}, i.e.
\be
{\vec B}^W\equiv W^{-1}\, {\vec B}\, W= -\frac{1}{e}\,\frac{{\hat r}}{r^2} \; T_R
\lab{secongmagfield}
\ee
Note that contrary to \rf{wuyangmagfield}, which is non-abelian and has three components, i.e. ${\vec B}={\vec B}_a\,T_a$, the magnetic field \rf{secongmagfield} is abelian and has just one component along the Lie algebra element $T_R$, defined by the direction of the unit vector ${\hat r}$, at the reference point $x_R$. The fact that ${\vec B}^W$ depends upon the choice of reference point does not introduce physical problems since the integral Yang-Mills equations transform in a covariant way under the change of that choice (see \cite{ym1} for details on that).

In the calculations involving  the Yang-Mills integral equations  we shall consider  purely spatial three-volumes $\Omega$, and so all derivatives of the time coordinate $x^0$, w.r.t. the parameters $\sigma$, $\tau$ and $\zeta$ vanish. Therefore, the components $\calb_{0i}$ and $A_0$, of the antisymmetric tensor and of the gauge field, will not enter in the calculation (in fact $A_0=0$ for the solution \rf{wuyangsol}). Since ${\widetilde F}_{ij}=0$ for the Wu-Yang monopole \rf{wuyangsol}, one has, from \rf{calbdef}, that $\calb_{ij}=i\,e\,\alpha\,F_{ij}=-i\,e\,\alpha\,\ve_{ijk}\,B_k$. Therefore, the connection $T\(\calb,A,\tau\)$ in \rf{eqforvb} becomes abelian, i.e. $T\(\calb,A,\tau\)=-i\,e\,\alpha\,\int_{0}^{2\,\pi}d\sigma\; B_i^W\,\ve_{ijk}\, \frac{d\,x^{j}}{d\,\sigma}\,\frac{d\,x^{k}}{d\,\tau}$. Consequently, the integration to obtain $V$ in \rf{eqforvb} does not need the surface ordering because the integrand in the definition of $T\(\calb,A,\tau\)$ lies in the direction of $T_R$ for any value of $\sigma$ and $\tau$. So one has that
\br
V\(\partial\Omega\) = \exp\left[-i\,e\,\alpha\,\int_{\partial\Omega}\,
{\vec B}^W\cdot d{\vec \Sigma}\right]
=\exp\left[i\,\alpha\, T_R\, \int_{\partial\Omega} d{\vec \omega}\cdot {\hat r}\right]
\lab{vdomegafinal}
\er
where we have denoted $\ve_{ijk}\, \frac{d\,x^{j}}{d\,\sigma}\,\frac{d\,x^{k}}{d\,\tau}\,d\sigma\,d\tau\,\equiv d\Sigma_i\equiv r^2\,d\omega_i$, with $d{\vec \Sigma}$ being the vector perpendicular to the surface $\partial\Omega$ at any given point, and whose modulus is the area element of $\partial\Omega$ at that point. On its turn $d{\vec \omega}$ is a vector also perpendicular to the surface $\partial \Omega$ at any given point, such that the scalar product $d{\vec \omega}\cdot {\hat r}$ is the solid angle element at that point of $\partial\Omega$ seen from the origin of the coordinate system. 
 We shall assume that the scanning of the surface $\partial\Omega$ by loops, parameterized by $\sigma$ and $\tau$, is such that $d{\vec \Sigma}$ points outward $\partial\Omega$. If that is not the case one gets a minus sign in the definitions of $d{\vec \Sigma}$ and $d{\vec \omega}$. 
 
 For the surface $\partial\Omega$ given in Figure \ref{fig:surfaceV} one has that
\be
V\(S_R^2\)=\exp\left[i\,4\,\pi\,\alpha\, T_R\right]\qquad\qquad {\rm and}\qquad\qquad 
V\(S_0^2\)=\exp\left[-i\,4\,\pi\,\alpha\, T_R\right]
\ee
since $d{\vec \Sigma}$ points outward $\partial\Omega$, and so outward $S_R^2$, and inward $S_0^2$. The quantity $V$ evaluated on the thin radial cylinder becomes $V\({\rm cylinder}\)=\one$, since $d{\vec \Sigma}\cdot {\hat r}=0$ on that surface because $d{\vec \Sigma}$ is perpendicular to the radial direction there. Consequently
\be
V\(\partial\Omega\) =V\(S_R^2\)\,V\({\rm cylinder}\)\,V\(S_0^2\)=\one
\lab{trivialsurfacestokes}
\ee
Such a result remains valid for any volume that does not contain the origin, since the r.h.s. of \rf{vdomegafinal} measures the solid angle associated to the surface $\partial\Omega$ as seen from the origin.   
 
 Another consequence of \rf{fconjugated} is that the commutator term in \rf{calkdef} (or \rf{calkym})  vanishes, i.e. 
\br
\sbr{\calb_{ij}^W\(\sigma^{\prime}\)-ie F_{ij}^W\(\sigma^{\prime}\)}{\calb_{kl}^W\(\sigma\)}=
-e^2\,\alpha\(\alpha-1\)\sbr{ F_{ij}^W\(\sigma^{\prime}\)}{F_{kl}^W\(\sigma\)}=0
\er
Note from \rf{calkym}  that such a  commutator term  contributes to the volume integral in \rf{stokes}, together with  ${\tilde j}_{\mu\nu\lambda}$ and ${\tilde J}_{\mu\nu\lambda}$, as densities  for the electric and magnetic charges. It is a non-linear term that corresponds in fact to the charges carried by the non-abelian gauge fields. In the case of the 't Hooft-Polyakov magnetic monopole such commutator term does not vanish and it accounts for the total of the dynamical magnetic charge of the configuration, since there are no magnetic sources associated to that solution \cite{directtest}. So, contrary to the 't Hooft-Polyakov monopole, the Wu-Yang monopole does not receive any contribution to its magnetic charge from such non-linear term, and so from the self-interacting non-abelian gauge fields.  

Using \rf{diffymeqs} one gets that
\br
\(D_i\,F_{jk}+D_j\,F_{ki}+D_k\,F_{ij}\)\,\frac{d\,x^{i}}{d\,\sigma}\,\frac{d\,x^{j}}{d\,\tau}\,
\frac{d\,x^{k}}{d\,\zeta} 
=j^0\,\ve_{ijk}\,\frac{d\,x^{i}}{d\,\sigma}\,\frac{d\,x^{j}}{d\,\tau}\,\frac{d\,x^{k}}{d\,\zeta}.
\er
Therefore, the quantity ${\cal K}$ introduced in \rf{calkdef}, for the Wu-Yang monopole and for the case of $\Omega$ being purely spatial,  becomes
\be
{\cal K}=ie\alpha
\int_{0}^{2\pi} \int_{0}^{2\pi}d\tau d\sigma V\,W^{-1}\,{j^0}\,W\,V^{-1}
\ve_{ijk}\,\frac{d\,x^{i}}{d\,\sigma}\,\frac{d\,x^{j}}{d\,\tau}\,\frac{d\,x^{k}}{d\,\zeta}.
\lab{calkj0}
\ee
We have seen in section \ref{sec:firstmagfield} that the density of magnetic charge $j_0$, for the Wu-Yang monopole, vanishes everywhere except perhaps at the origin. Since we are dealing with volumes $\Omega$ that do not contain the origin, one then concludes that ${\cal K}=0$, and so 
\be
U\(\Omega\)= \one \; ;\qquad\qquad\qquad \mbox{\rm for volumes $\Omega$ not containing the origin}
\lab{trivialvolumestokes}
\ee
From  \rf{trivialsurfacestokes} and \rf{trivialvolumestokes} one then observes that the surface and volume integrals agree when the volume $\Omega$ does not contain the origin, and so the integral Yang-Mills equation \rf{stokes} work perfectly well for the Wu-Yang monopole in that situation. 

The question now is if one can extend the application of the integral Yang-Mills equations to volumes 
${\hat \Omega}$ containing the origin. In order to elucidate that, let us apply distribution theory to analyze the  divergence of the magnetic field  \rf{secongmagfield}. Note that each component of the magnetic field \rf{secongmagfield} has a singularity $1/r^2$ at the origin, and so they are locally summable functions in $\IR^3$. Therefore, similarly to \rf{distributionbi}, one can introduce the distributions $\langle T_{B_i^W},\Phi\rangle \equiv \int d^3 x\, B_i^W\, \Phi$. Following \rf{graddef} we have that the distribution associated to the divergence of ${\vec B}^W$ is
\br
\langle T_{{\vec \nabla}\cdot {\vec B}^W},\Phi\rangle &=&\int d^3{\vec r}\; {\vec \nabla}\cdot {\vec B}^W\; \Phi 
= - \int d^3{\vec r}\,  {\vec B}^W\cdot {\vec \nabla}\Phi
\nonumber\\
&=& \frac{1}{e}\,\int d\theta\,d\vp\; \sin\theta\;T_R\;\left[\Phi\(\infty\,,\theta\,,\,\vp\)-\Phi\(0\,,\theta\,,\,\vp\)\right]
\nonumber\\
&=&-\frac{4\,\pi}{e}\, T_R\, \Phi\({\vec 0}\)
\lab{graddefw}
\er
We now have to find a localy summable function ${\hat u}$ such that $\int d^3{\vec r}\; {\hat u}\; \Phi$ gives that same result for any test function $\Phi$, and so we can say that ${\vec \nabla}\cdot {\vec B}^W={\hat u}$. There are however two functions satisfying that condition given by 
\be
{\hat u}_1= -\frac{1}{e}\, \frac{\delta\(r\)}{r^2}\, T_R\;;\qquad\qquad\qquad 
{\hat u}_2= -\frac{4\,\pi}{e}\, \delta^{(3)}\({\vec r}\)\, T_R
\lab{twopossibleus}
\ee
where $\delta^{(3)}\({\vec r}\)$ is the three-dimensional Dirac delta function centered at the origin, and $\frac{\delta\(r\)}{r^2}$ is its radial part. Indeed, using the same reasonings leading to \rf{checkgeneralu}, we have  that
\br
\int d^3{\vec r}\; {\hat u}_1\; \Phi&=& -\frac{1}{e}\, T_R\, \int d^3{\vec r}\; \frac{\delta\(r\)}{r^2}\, \Phi
=-\frac{4\,\pi}{e}\, T_R\, \Phi\({\vec 0}\)
\lab{hatu1distribution}
\er
and 
\br
\int d^3{\vec r}\; {\hat u}_2\; \Phi&=& -\frac{4\,\pi}{e}\, T_R\, \int d^3{\vec r}\; \delta^{(3)}\({\vec r}\)\, \Phi
=-\frac{4\,\pi}{e}\, T_R\, \Phi\({\vec 0}\)
\er

 Consider now a Lie algebra valued quantity $X\(x_P\)$ evaluated on a given point $x_P$ of the volume $\Omega$, and consider the conjugation 
$W^{-1}\,X\(x_P\)\,W$, where the Wilson line operator $W$ is obtained by integrating \rf{eqforw} from the reference point $x_R$ to the point $x_P$ through the loop to which $x_P$ belongs to in that given scanning. Therefore, using \rf{eqforw}, and performing calculations similar to those in \rf{rtconstant}, one gets that 
\be
\frac{d\;}{d\sigma} \(W^{-1}\,X\(x_P\)\,W\)=W^{-1}\, D_i\, X\(x_P\)\, W\, \frac{d\,x^i}{d\,\sigma}
\lab{xconjugatedderivaative}
\ee
and so we have that
\be
\left[\partial_i\(W^{-1}\,X\(x_P\)\,W\)-W^{-1}\, D_i\, X\(x_P\)\, W\right]\, \frac{d\,x^i}{d\,\sigma}=0
\ee
Therefore, the component of $\left[\partial_i\(W^{-1}\,X\(x_P\)\,W\)-W^{-1}\, D_i\, X\(x_P\)\, W\right]$, tangent to the curve, where $W$ is evaluated, at the point $x_P$  has to vanish. If one takes $x_P$ to be the origin of the Cartesian coordinates, it turns out that all tangent vectors to the curve  are radial, and so one must have
\be
\partial_r\(W^{-1}\,X\( 0\)\,W\)-W^{-1}\, D_r\, X\( 0\)\,W=0
\lab{niceradialidentity}
\ee
where $r$ is the radial variable in the spherical polar coordinates. As shown in the appendix of \cite{ym1}, there are no problems in calculating the  Wilson line $W$ on curves passing through the origin for the Wu-Yang connection $A_i$, given in \rf{wuyangsol}, due to its singularity. Therefore, $W$ in \rf{niceradialidentity} is well defined even when integrated from the reference point $x_R$ up to the origin.  Taking $X$, in \rf{niceradialidentity},  to be the radial component of the magnetic field one gets
\be
\partial_r\(W^{-1}\,B_r\( 0\)\,W\)-W^{-1}\, D_r\, B_r\( 0\)\,W=0
\lab{niceradialidentityforb}
\ee
Since the Wu-Yang magnetic field \rf{wuyangmagfield} has only radial components, one can write \rf{niceradialidentityforb} as 
\be
{\vec \nabla}\cdot {\vec B}^W\( 0\)-W^{-1}\, {\vec D}\cdot {\vec B}\( 0\)\,W={\vec \nabla} \cdot{\vec B}^W\( 0\)-W^{-1}\, {\vec \nabla}\cdot {\vec B}\( 0\)\,W=0
\lab{niceradialidentityforbtotal}
\ee
where in the first equality we have used \rf{nullcommutatorab}, and where ${\vec B}^W$ is defined in \rf{secongmagfield}. We have shown that the divergences of ${\vec B}$ and ${\vec B}^W$, for the Wu-Yang solution,  vanish everywhere in $\IR^3$, except perhaps at the origin of the Cartesian coordinates.  Therefore, we have established that everywhere in $\IR^3$ the following relation holds true for the Wu-Yang monopole solution  
\be
{\vec \nabla} \cdot{\vec B}^W=W^{-1}\, {\vec \nabla}\cdot{\vec B}\,W
\lab{finalrelationbbw}
\ee
Therefore, using \rf{generalu} and \rf{fconjugated} we have that
\be
W^{-1}\,{\vec \nabla}\cdot {\vec B}\,W=C\, W^{-1}\,{\hat r}\cdot {\vec T}\,W\,\frac{\delta\(r\)}{r^2}
= C\, \frac{\delta\(r\)}{r^2}\,T_R
\lab{conjugateddivergenceb}
\ee
Note that the conjugation by the Wilson line $W$ has mapped the function \rf{generalu}, that has components along the three generators of the $SU(2)$ Lie algebra, $T_a$, $a=1,2,3$, to a given  function that has only one component along the generator $T_R$ of the same Lie algebra. That is a very important fact since, as shown in \rf{divb=0} and \rf{checkgeneralu}, it is the term ${\hat r}\cdot {\vec T}$, and not the delta function, that makes the integral of those functions to vanish when multiplied by any test function, and consequently making the associated distribution equivalent to the zero distribution. On the other hand, the function on the r.h.s. of \rf{conjugateddivergenceb}, or equivalently the function ${\hat u}_1$ given in \rf{twopossibleus}, do not lead to the zero distribution, as shown in \rf{hatu1distribution}, because it does not have that angular term ${\hat r}\cdot {\vec T}$.  

Consequently, comparing  \rf{finalrelationbbw} and \rf{conjugateddivergenceb} with \rf{twopossibleus} one concludes that the function ${\hat u}_2$ in  \rf{twopossibleus}  has to be discarded, and the value of $C$ in \rf{generalu} has to be fixed as 
\be
C=-\frac{1}{e}
\lab{goodc}
\ee
and so we have that
\be
{\vec \nabla} \cdot{\vec B}^W= -\frac{1}{e}\, \frac{\delta\(r\)}{r^2}\, T_R\;;\qquad\qquad\qquad {\rm and}\qquad\qquad\qquad {\vec \nabla}\cdot{\vec B}= -j_0=-\frac{1}{e}\,{\hat r}\cdot {\vec T}\,\frac{\delta\(r\)}{r^2}
\lab{finaldivergences}
\ee
where we have used the relation \rf{j0divegencerel} between the divergence of $B_i$ and the density of magnetic charge $j_0$. 
Using \rf{nullcommutatorab} one then concludes that \rf{finaldivergences} justifies our claim \rf{result}. 

We stress that in the local analysis of the Yang-Mills differential equations,  discussed in section \ref{sec:firstmagfield}, we obtained that all the locally summable functions given in \rf{generalu} lead to the zero distribution.  However, functions that lead to the same distribution  can differ in a set of zero measure \cite{distribution}, which in our case is the origin of the Cartesian coordinate system. Such a difference, in a set of zero measure, is of crucial importance when analyzing the global aspects of the Wu-Yang solution in the context of the integral Yang-Mills equations. The equation   \rf{finalrelationbbw}, which  is a non-local relation due to the Wilson line $W$,  is responsible for the selection of the unique function,  given in  \rf{result} and \rf{finaldivergences},  that is compatible with the local and global aspects of the Wu-Yang solution.

We now turn to the the integral Yang-Mills equations and apply them to volumes ${\hat \Omega}$ that contain the origin inside it. 
Clearly, the surface integral can be easily calculated using the same procedures leading to \rf{vdomegafinal}, and the result is that the flux of the magnetic field \rf{secongmagfield} is non-trivial, since the l.h.s. of  \rf{vdomegafinal} measures the solid angle, seen from the origin, associated to surface $\partial {\hat \Omega}$ (border of ${\hat \Omega}$). The result is 
\be
V\(\partial{\hat \Omega}\)= \exp\left[i\,4\,\pi\,\alpha\,T_R\right]\;; \qquad\qquad\qquad\mbox{\rm for a volume ${\hat \Omega}$ containing the origin}
\lab{partialomegaorigin}
\ee
Using  \rf{fconjugated} and \rf{finaldivergences} one gets that $W^{-1}\,j_0\,W=\frac{1}{e}\,\frac{\delta\(r\)}{r^2}\, T_R$.  For the Wu-Yang monopole we have that ${\tilde F}_{ij}=0$ (no electric field). Therefore, from \rf{fconjugated} and \rf{calbdef} one gets that $T\(\calb,A,\tau\)$ given in \rf{eqforvb} becomes 
\be
T\(\calb,A,\tau\)=\frac{1}{e}\, T_R\,
\int_{0}^{2\,\pi}d\sigma\; \ve_{ijk}\,\frac{x^k}{r^3}\, \frac{d\,x^{i}}{d\,\sigma}\,\frac{d\,x^{j}}{d\,\tau} 
\ee
Consequently, the integration of the equation for $V$ in \rf{eqforvb} leads to operators $V$ belonging to the $U(1)$ group generated by $T_R$, and so $V\,W^{-1}\,j_0\,W\,V^{-1}= \frac{1}{e}\, \frac{\delta\(r\)}{r^2}\, T_R$. From \rf{calkj0} one observes that ${\cal K}$ has components in the direction of $T_R$ only, and so the integration in \rf{eqforv2}, to obtain $U$, does not need the volume ordering, since ${\cal K}$ becomes abelian. Therefore, considering a volume ${\hat \Omega}$ containing the origin, one gets 
\be
U\({\hat \Omega}\)= \exp\left[i\,\alpha\,T_R\int_{{\hat \Omega}} \, d^3 {\vec r}\, \frac{\delta\(r\)}{r^2}\right]
= \exp\left[i\,4\,\pi\,\alpha\,T_R\right]
\lab{omegaorigin}
\ee
where we have denoted $d^3{\vec r}=\ve_{ijk}\,\frac{d\,x^{i}}{d\,\sigma}\,\frac{d\,x^{j}}{d\,\tau}\,\frac{d\,x^{k}}{d\,\zeta}\,d\sigma\,d\tau\,d\zeta$. We have assumed that the scanning of the surfaces with loops, parameterized by $\sigma$ and $\tau$, is such that the vector $\ve_{ijk}\,\frac{d\,x^{i}}{d\,\sigma}\,\frac{d\,x^{j}}{d\,\tau}$ points outwards the surface. Since 
$\zeta$ grows in the direction outwards the surface, it turns out that $d\zeta\,d\tau\,d\sigma\,\ve_{ijk}\,\frac{d\,x^{i}}{d\,\sigma}\,\frac{d\,x^{j}}{d\,\tau}\,\frac{d\,x^{k}}{d\,\zeta}=d^3{\vec r}$ (if the scanning does not satisfy that condition one gets $-d^3{\vec r}$ instead). Comparing \rf{partialomegaorigin} and \rf{omegaorigin} one concludes that, as a consequence of \rf{finaldivergences},  the Wu-Yang monopole configuration also satisfies the Yang-Mills integral equations   for volumes ${\hat \Omega}$  containing the origin. Note that the agreement between  \rf{partialomegaorigin} and \rf{omegaorigin} occurs only for the functions given in \rf{finaldivergences}. Therefore, the integral  Yang-Mills equations also select, among the functions \rf{generalu},  the unique function corresponding to the value of $C$ given in \rf{goodc}.

In the integral Yang-Mills equations the components of the field tensor (and its dual) always appear conjugated by the Wilson line $W$. For the case of the Wu-Yang monopole configuration we have just seen that those conjugated terms are all in the direction of the Lie algebra element $T_R$, the value of ${\hat r}\cdot {\vec T}$ at the reference point $x_R$. Therefore, everything becomes abelian and the surface and volume orderings are unnecessary. In addition, since the parameter $\alpha$ appearing in the integral equations is arbitrary (see \cite{directtest}),    then one can expand the integral Yang-Mills equations in powers of that parameter. The linear term in $\alpha$, for the Wu-Yang monopole configuration,  gives the integral equation, valid for any volume ${\hat \Omega}$ including or not the origin,   
\be
\int_{\partial{\hat \Omega}}\,{\vec B}^W\cdot d{\vec \Sigma}= 
\int_{{\hat \Omega}} d^3 {\vec r}\;\; W^{-1}\,{\vec \nabla}\cdot {\vec B}\,W= 
\int_{{\hat \Omega}} d^3 {\vec r}\;\; {\vec \nabla}\cdot {\vec B}^W
\lab{abelianstokesym}
\ee
where in the last equality we have used \rf{finalrelationbbw}. But that is exactly the Gauss law (abelian Stokes theorem) for the magnetic field ${\vec B}^W$, introduced in \rf{secongmagfield}, and it therefore follows from the integral Yang-Mills equation \rf{stokes} for the case of the Wu-Yang monopole solution. 

The same reasonings apply to the conserved charges which are the eigenvalues of the operator in \rf{chargeym}. In fact, \rf{chargeym} have the same structure of the integral Yang-Mills equations, and so for the Wu-Yang monopole all terms lie in the direction of the same Lie algebra element $T_R$. Therefore, expanding both sides \rf{chargeym} in powers of $\alpha$ one gets conserved charges for every term of the series. However, due to the abelian structure the higher charge operators are powers of the linear one and one gets just a finite number of conserved magnetic charges. They are the eigenvalues of the following operator, coming from the linear term in $\alpha$ of the expansion  of \rf{chargeym},   
\be
Q_{YM}\equiv \int_{\partial {\cal S}}\,{\vec B}^W\cdot d{\vec \Sigma}= 
\int_{{\cal S}} d^3 {\vec r}\;\; {\vec \nabla}\cdot {\vec B}^W= -\frac{4\,\pi}{e}\, T_R
\lab{truemagcharge}
\ee
where ${\cal S}$ is the spatial sub-manifold, i.e. ${\cal S}\equiv \IR^3$. Contrary to the usual Noether charges \rf{magchargeallspace}, these are non-vanishing and invariant under general gauge transformations \cite{ym1,ym2}. In addition, they do not depend upon the choice of the reference point $x_R$, since as shown in \cite{ym1,ym2}, $T_R$ changes by conjugation by the Wilson line operator $W$,  joining the new and old reference points through a given curve, and so its eigenvalues do not change. In a spin $j$ representation of the $SU(2)$ gauge group, the magnetic charges are 
\be
\mbox{\rm eigenvalues of $Q_{YM}$}=-\frac{4\,\pi}{e}\, \(j\,,\,j-1\,,\, \ldots \,,\,-j\)
\ee
Therefore we have  established that the Wu-Yang monopole solution does possess  dynamically conserved non-vanishing magnetic charge associated to it.  
As discussed in \cite{ym1,ym2}, at the classical level there are no ways of fixing the representation where the eigenvalues should be evaluated. A more detailed discussion of that issue is given in \cite{remark}. 
 
We would like to conclude this section with two comments concerning the role of the singularity of the Wu-Yang monopole in two other calculations. First, note that the gauge potential $A_i$ given in \rf{wuyangsol}  is also singular at the origin, and one could wonder if the evaluation of the curl of $A_i$, to obtain $B_i$, leads to singular terms like a Dirac string. However using  distribution theory to evaluate the curl as $\int d^3{\vec r}\, {\vec \nabla}\wedge {\vec A}\; \Phi = - \int d^3{\vec r}\,  {\vec A}\wedge {\vec \nabla}\Phi$, one gets that ${\vec \nabla}\wedge {\vec A}=\frac{2}{e}\,\frac{{\hat r}}{r^2}\,{\hat r}\cdot {\vec T}$, and so there are no singularities besides $1/r^2$. 

Second, for the Wu-Yang monopole solution \rf{wuyangsol} the electric current $J^{\mu}$, given in  \rf{diffymeqs}, has to vanish. Indeed, the time component $J^0$ vanishes because the solution is static, $A_0=0$, and $F_{0i}=0$. For the space components $J^i$ one has to take care in the evaluation of $\partial_j F^{ji}$, because of the singularity at the origin. However, using distribution theory as we did above one can show that  $\partial_j F_{ji}= \frac{1}{e}\,\ve_{ijk}\,\frac{x^j}{r^4}\,T_k$, and so there is only the usual singularity at the origin $1/r^3$, and nothing else. Then one obtains that $J^i$ has indeed to vanish due to the differential Yang-Mills equations.

\section{Conclusions}
\label{sec:conclusions}
\setcounter{equation}{0}

We have shown that the Wu-Yang monopole solution does need a source to sustain it. That was established using distribution theory to the differential Yang-Mills equations and also the integral Yang-Mills equations constructed in \cite{ym1,ym2}. In fact, these integral equations played a crucial role in the analysis because it introduced another type of magnetic field, introduced in \rf{secongmagfield}, that allowed to fix in a unique way the function that characterizes the source of Wu-Yang magnetic field.

It is well know  in distribution theory that, if two locally summable functions give the same result when multiplied by any test function $\Phi$ and integrated in the domain of these tests function, then those two functions lead to the same distribution. Therefore, in this sense, all the functions \rf{generalu} lead to the zero distribution. Consequently, the point source introduced in this paper (see \rf{result}), to sustain the Wu-Yang solution, does not violate the Bianchi identities in the sense of distribution theory, since the distribution associated to such a source is the zero distribution. That is why we call it a mild source. However, functions leading to the same distribution  can  differ in a set of zero measure, which is exactly what happens with the functions   \rf{generalu}. Such a set of zero measure (the origin) is crucial for the integral Yang-Mills  equations, and also for the second magnetic field introduced in \rf{secongmagfield}. The local analysis of the Yang-Mills differential equations, through distribution theory, does not fix the source needed for the Wu-Yang monopole solution, since any of the functions \rf{generalu} lead to the same distribution.  The integral Yang-Mills equations bring the global properties of the solution into the analysis and the differences among those sources, in a set of zero measure, matters in a crucial way. In fact, as we have shown, the integral equations are compatible with only one of those sources, namely that one given on the r.h.s. of \rf{result}.  The selection of that unique source is a consequence of the relation \rf{finalrelationbbw} between the divergences of the two types of magnetic fields relevant to the physical properties of the Wu-Yang monopole solution. Note that \rf{finalrelationbbw} is a non-local relation since it involves the Wilson line $W$.

Finally we would like to comment that the $SU(2)$ Wu-Yang monopole solution can be embedded in any Yang-Mills theory associated to any compact group $G$ that contains $SU(2)$ as a subgroup. Therefore, the Wu-Yang monopole is a classical solution of all those Yang-Mills theories where the gauge symmetry is not broken. In particular it is a classical solution of $QCD$, and the results of the present paper might have important physical consequences there. It would be interesting to investigate if the  point magnetic source constructed here has any physical role in QCD.

\vspace{2cm}

\noindent {\bf Acknowledgments:} We are very grateful  to Profs. Sergio Lu\'is Zani and Farid  Tari for many helpful discussions on distribution theory. We are also grateful to J.C. Barata, R. Koberle, E.C. Marino, J. Sanch\'ez Guill\'en,  Yasha Shnir, P. Sutcliffe, P. A. F. da Veiga and W.  Zakrzewski for valuable comments on the manuscript. The authors are grateful to partial finantial support by FAPES under contract number  0447/2015. LAF is partially supported by CNPq-Brazil.

\newpage

\appendix

\section{The Yang-Mills integral equations}
\label{app:stokes}
\setcounter{equation}{0}

The basic ingredient for the construction of integral equations for gauge theories is the Stokes theorem for an antisymmetric rank two tensor $\mathcal{B}_{\mu\nu}$, that relates the flux of such a tensor through a surface $\partial \Omega$, which is the border of a volume $\Omega$, to the integral of the curvature of that tensor on the volume $\Omega$. For an abelian gauge theory, like electrodynamics,  one uses the well known abelian Stokes theorem
\be
\int_{\partial \Omega}  \mathcal{B}_{\mu\nu}\;dx^\mu\wedge dx^\nu = \int_{\Omega} \left[ \partial_{\mu} \mathcal{B}_{\nu\rho}+\partial_{\nu} \mathcal{B}_{\rho\mu}+\partial_{\rho} \mathcal{B}_{\mu\nu}\right]dx^\mu\wedge dx^\nu\wedge dx^{\rho}
\lab{abelstokes}
\ee
Indeed, by taking $\mathcal{B}_{\mu\nu}$ to be a linear combination of the field tensor $F_{\mu\nu}$, and its Hodge dual, $\widetilde{F}_{\mu\nu}\equiv \frac{1}{2}\,\ve_{\mu\nu\rho\sigma}\,F^{\rho\sigma}$, i.e.
\be
\mathcal{B}_{\mu\nu}\equiv \alpha F_{\mu\nu}+\beta \widetilde{F}_{\mu\nu}
\lab{bfrel}
\ee
and using Maxwell's equations, $\partial_{\mu} F_{\nu\rho}+\partial_{\nu} F_{\rho\mu}+\partial_{\rho} F_{\mu\nu}=0$, and $\partial_{\mu} {\widetilde F}_{\nu\rho}+\partial_{\nu} {\widetilde F}_{\rho\mu}+\partial_{\rho} {\widetilde F}_{\mu\nu}={\widetilde J}_{\mu\nu\rho}$, where  ${\widetilde J}_{\mu\nu\rho}\equiv \frac{1}{3!}\ve_{\mu\nu\rho\sigma}\,J^{\sigma}$, is the Hodge  dual of the  electric four-current $J^{\mu}$, one gets from \rf{abelstokes} the integral equations for electrodynamics, i.e. 
\be
\int_{\partial \Omega} \left[\alpha F_{\mu\nu}+\beta \widetilde{F}_{\mu\nu}\right]\;dx^\mu\wedge dx^\nu = \int_{\Omega} \;\beta\,{\widetilde J}_{\mu\nu\rho}\;dx^\mu\wedge dx^\nu\wedge dx^{\rho}
\ee
In the case where the volume $\Omega$ is purely spatial, one gets for $\alpha=0$ the Gauss law, $\int_{\partial\Omega} {\vec E}\cdot d{\vec \Sigma} = q$, with $q$ the electric charge inside $\Omega$, and for $\beta=0$ one gets $\int_{\partial\Omega} {\vec B}\cdot d{\vec \Sigma} = 0$, where ${\vec E}$ and ${\vec B}$ are the electric and magnetic fields respectively, i.e. $E_i=F_{0i}$ and $B_i=-\frac{1}{2}\ve_{ijk}F_{jk}$. For the case where $\Omega$ is a hyper-cylinder with the bottom and top bases being purely spatial and the height purely temporal, one gets the integral Faraday law for $\beta=0$ (in the limit where the cylinder's height goes to zero), and its electric counterpart for $\alpha=0$.  

The construction of the integral equations for non-abelian gauge theories requires the non-abelian version of the Stokes theorem for an antisymmetric rank two tensor, and  that has been constructed in \cite{afg1,afg2,ym1,ym2}. The main difficulty in the implementation of that theorem is that  the components of that tensor at different points of space-time do not commute, and so the integrals have to be ordered. The best way of doing that is to scan the volume $\Omega$ with closed two dimensional surfaces based at a reference point $x_R$ on the border $\partial \Omega$ of $\Omega$. We label such surfaces with a parameter $\zeta$ such that $\zeta=0$ corresponds to the infinitesimal closed surface around $x_R$, and $\zeta=2\,\pi$ to the border $\partial\Omega$. Each closed surface on its turn is scanned by loops starting and ending at $x_R$, and they are labelled by a parameter $\tau$, varying from $0$ to $2\,\pi$, such that $\tau=0$ corresponds to a infinitesimal loop around $x_R$, then as $\tau$ varies the loops scan the surface and it ends, at $\tau=2\,\pi$, on another infinitesimal loop around $x_R$ on the other side of it. The loops are parameterized by a parameter $\sigma$ such that $\sigma=0$ and $\sigma=2\,\pi$ correspond to the starting and ending points of the loop that coincide with the reference point $x_R$ on $\partial \Omega$.

Another difficulty related to the non-commutativity of the components of the tensor $\mathcal{B}_{\mu\nu}$ concerns the gauge covariance of the non-abelian Stokes theorem. If the the rank two tensor is going to be related to the  field tensor of the non-abelian gauge theory as in \rf{bfrel}, then under a gauge transformation $A_{\mu}\rightarrow g\,A_{\mu}\,g^{-1}+\frac{i}{e}\,\partial_{\mu}g\,g^{-1}$, where $A_{\mu}$ is the non-abelian gauge field, and $e$ the gauge coupling constant, one gets that $\mathcal{B}_{\mu\nu}$ transforms as $\mathcal{B}_{\mu\nu}\rightarrow g\,\mathcal{B}_{\mu\nu}\,g^{-1}$. But any surface ordered integral of such a tensor will transform under the gauge group in a terrible way. To circumvent it we conjugate $\mathcal{B}_{\mu\nu}$ with the Wilson line $W$ defined in  \rf{eqforw}. Under a gauge transformation, $W$ transforms as $W\rightarrow g\(x_f\)\,W\,g^{-1}\(x_i\)$, where $x_i$ and $x_f$ are respectively the initial and final points of the curve where $W$ is defined. In a given scanning any point of the volume $\Omega$ belongs to a unique closed surface and to a unique loop. Therefore, for any point $x$ of the volume $\Omega$ we consider a Wilson line $W\(x\)$ obtained by integrating \rf{eqforw} from the reference point $x_R$ up to $x$ along the loop to which $x$ belongs in the chosen scanning of $\Omega$. Therefore, the quantity $\mathcal{B}_{\mu\nu}^W\equiv W^{-1}\(x\)\,\mathcal{B}_{\mu\nu}\(x\)\,W\(x\)$, transforms as $\mathcal{B}_{\mu\nu}^W\rightarrow g\(x_R\)\, \mathcal{B}_{\mu\nu}^W\,g^{-1}\(x_R\)$. The ordered surface integral of such quantity has a quite simple gauge transformation, since its transformation law only involves the gauge group element evaluated at the reference point $x_R$, and so effectively it transforms under a global gauge transformation.  

The first step to construct the non-abelian Stokes theorem is to associate to every closed surface scanning the volume $\Omega$,  a quantity $V$ as defined in \rf{eqforvb}. Note that $V$ is a surface  ordered integral, and the ordering is given by the chosen scanning of $\Omega$ as explained above. Indeed, we first evaluate $T\(\calb,A,\tau\)$ on each loop scanning the surface, and then integrate along the surface according to the ordering of the loops labelled by $\tau$. The second step is to determine how $V$ changes when one performs an infinitesimal variation of the closed surface associate to it. So, each point $x^{\mu}$ of the surface is changed as $x^{\mu}\rightarrow x^{\mu}+\delta x^{\mu}$, and the variation $\delta x^{\mu}$ is perpendicular to the surface. However, the reference point $x_R$ is kept fixed, i.e. $\delta x^{\mu}\(x_R\)=0$. The quantity $V$ changes to $V+\delta V$, and such variation is given by  \cite{afg1,afg2,ym1,ym2}
\br
\delta V\,V^{-1}&=&
\int_0^{2\,\pi}d\tau \,V\(\tau\)\int_0^{2\,\pi}d\sigma \,\left\{ 
 W^{-1}\,
\left[D_{\lambda}\calb_{\mu\nu}+D_{\mu}\calb_{\nu\lambda}+D_{\nu}\calb_{\lambda\mu}\right]
\,W\frac{d\,x^{\mu}}{d\,\sigma}\,\frac{d\,x^{\nu}}{d\,\tau}\,
\delta\,x^{\lambda}\right.\nonumber\\
&-&\left. \int_0^{\sigma}d\sigma^{\prime}
\sbr{\calb_{\kappa\rho}^W\(\sigma^{\prime}\)-ie F_{\kappa\rho}^W\(\sigma^{\prime}\)}
{\calb_{\mu\nu}^W\(\sigma\)}\frac{dx^{\kappa}}{d\sigma^{\prime}}\frac{dx^{\mu}}{d\sigma}\right.
\nonumber\\
&&\left.\times
\(\frac{d\,x^{\rho}\(\sigma^{\prime}\)}{d\,\tau}\delta\,x^{\nu}\(\sigma\)
-\delta\,x^{\rho}\(\sigma^{\prime}\)\,\frac{d\,x^{\nu}\(\sigma\)}{d\,\tau}\)\right\} V^{-1}\(\tau\)
\lab{varofv}
\er  
The third step to obtain the non-abelian Stokes theorem is to realize that as we vary the parameter $\zeta$, used in the scanning of $\Omega$ as discussed above, we vary the surfaces scanning it. Therefore, the relation \rf{varofv} can be seen as a differential equation for $V$ in the parameter $\zeta$. Indeed, dividing both sides of  \rf{varofv} by $\delta \zeta$, and taking the limit $\delta \zeta\rightarrow 0$, one gets that  \rf{varofv} becomes the differential equation 
\be
\frac{d\, V}{d\,\zeta} - {\cal K}\, V=0
\lab{secondeqforv}
\ee
with ${\cal K}$ given by \rf{calkdef}. We now have two different ways of obtaining $V$ for a given closed surface. The first one is through \rf{eqforvb} and the other one through \rf{secondeqforv}. Integrating \rf{secondeqforv} from the infinitesimal closed surface around the reference point $x_R$, corresponding to $\zeta=0$, up to the closed surface which matches to the border $\partial \Omega$, corresponding to $\zeta=2\,\pi$, we obtain the quantity $V$ associated to the border of $\Omega$, i.e. $V\(\partial \Omega\)$. On the other hand, if we integrate  \rf{eqforvb} on $\partial \Omega$, we must obtain the same quantity. We then obtain the  non-abelian Stokes theorem as stated in \rf{stokes} where $U\(\Omega\)$ stands for  $V\(\partial \Omega\)$ as obtained from \rf{secondeqforv}, or equivalently \rf{eqforv2}.

The non-abelian Stokes theorem \rf{stokes} constitutes a generalization of the abelian theorem \rf {abelstokes}, in the sense that \rf {abelstokes} follows from \rf{stokes} in the case where the tensor $\mathcal{B}_{\mu\nu}$ and the connection $A_{\mu}$ lie on an abelian Lie algebra. In fact, the connection disappears from  the statement of the theorem since it only enters in it through the Wilson line which now is abelian and so its conjugation is trivial, and also through the commutator in the covariant derivatives $D_{\mu}$, which also trivializes. In addition, the ordering of the surface and volume integrals become unnecessary since everything is now abelian. It is worth noting that the non-abelian theorem \rf{stokes} is not only defined for a given volume and its border, but also for a given chosen scanning of the volume. By changing the scanning without changing the physical volume, both sides of \rf{stokes} do change. However, by the construction of the theorem it is guaranteed that both sides remain equal to each other in the new scanning. So, it is correct to say that the  non-abelian Stokes theorem \rf{stokes} transforms in a covariant way under repameterization of the volumes and surfaces. In fact the correct mathematical language for the theorem \rf{stokes} is that of generalized loop spaces. The scanning of the volume $\Omega$ make its points to be functions of the scanning parameters, i.e. $x^{\mu}=x^{\mu}\(\sigma, \tau ,\zeta\)$. The closed surfaces based at the reference point are in fact images of a map from a two-sphere $S^2$, with coordinates $\(\sigma, \tau\)$, and the relevant loop space is given by the following space  of functions $\gamma$ 
\be
{\cal L}^{(2)}\equiv \{ \gamma\, : S^2 \rightarrow \Omega\, \mid \mbox{\rm north pole of} \; S^2 \rightarrow x_R\}
\ee
Each closed surface based at $x_R$,  scanning $\Omega$, is a point of ${\cal L}^{(2)}$, and since the scanning makes the volume $\Omega$ to be a collection of such points, one can see $\Omega$ as a path in ${\cal L}^{(2)}$. Obviously, there is an infinite number of paths in  ${\cal L}^{(2)}$ corresponding to the same physical volume $\Omega$, and that is the reparameterization freedom one has in the formulation of \rf{stokes}. Therefore, \rf{stokes} is formulated on a given path in ${\cal L}^{(2)}$, and not only on the physical volume $\Omega$. 

The theorem \rf{stokes} also transforms in a covariant way under gauge transformations, since all quantities appear conjugated by the Wilson line and so by the arguments given above, both sides of \rf{stokes} transform by conjugation by the  element of the gauge group evaluated at the reference point, i.e. $g\(x_R\)$. Note in addition, that all contractions of indices appearing in \rf{stokes} are pure sums, and do not involve a metric. Therefore, the non-abelian Stokes theorem  \rf{stokes} is valid for volumes on a curved space-time. The only restriction is that the volumes must be topologically trivial, i.e. should have no holes or handles. It is not difficult to adapt the theorem to the cases where the volumes are topologically non-trivial.  

The integral equations for non-abelian gauge theories are a direct consequence of the non-abelian Stokes theorem \rf{stokes} and the differential Yang-Mills equations \rf{diffymeqs}, in the same way the integral equations  for electrodynamics are a direct consequence of the abelian Stokes theorem \rf{abelstokes} and Maxwell's equations. Indeed, replacing the rank two tensor $\mathcal{B}_{\mu\nu}$ in \rf{stokes} by the linear combination \rf{calbdef}, and making use of the differential Yang-Mills equations \rf{diffymeqs}, one gets the integral equations 
\be
 P_2 \, e^{i\,e\,\int_{\partial\Omega}d\tau d\sigma \,W^{-1}\,\left[\alpha\,F_{\mu\nu}+\beta\,{\widetilde F}_{\mu\nu}\right]\,W\, \frac{dx^{\mu}}{d\sigma}\,\frac{d\,x^{\nu}}{d\,\tau}}= P_3\, e^{\int_{\Omega} d\zeta \,{\cal K}}
 \lab{integralmeqs}
 \ee
with ${\cal K}$ being given now by \rf{calkym}. The relations \rf{integralmeqs} are the integral equations for Yang-Mills theories \cite{ym1,ym2}. 

The inverse relation is also true, i.e. the Yang-Mills differential equations follow from the integral equations \rf {integralmeqs}. Indeed, consider the limit where the volume $\Omega$ becomes an infinitesimal volume around the reference point $x_R$ of size $\ve$. By expanding both sides of  \rf {integralmeqs} in powers of $\ve$, the differential Yang-Mills equations are recovered as the terms in first order in $\ve$. To make it concrete, take $\Omega$ as the infinitesimal cube of sides $dx^\mu$, $dx^\nu$ and $dx^\lambda$, and $x_R$ to be located at one of its vertices, both sides of the integral equation \rf{integralmeqs} are written as a power series in the volume $d^3x\equiv \frac{1}{3!}\epsilon_{\mu\nu\lambda}dx^\mu \wedge dx^\nu \wedge dx^\lambda$ and then, performing such expansion, every term is Taylor expanded around the reference point, giving, to the first non-trivial order
\begin{eqnarray*}
 P_2 \, e^{i\,e\,\int_{\partial\Omega}d\tau d\sigma \,W^{-1}\,\left[\alpha\,F_{\mu\nu}+\beta\,{\widetilde F}_{\mu\nu}\right]\,W\, \frac{dx^{\mu}}{d\sigma}\,\frac{d\,x^{\nu}}{d\,\tau}} &\approx& \one + ie\left[ D_\lambda\left( \alpha F_{\mu\nu}+\beta \widetilde{F}_{\mu\nu} \right) 
 \right. \nonumber\\
 && \left. + \;\textrm{cyclic permutation}\right]\;dx^\mu dx^\nu dx^\lambda,\\
P_3\,e^{\int_\Omega \mathcal{K}d\zeta} &\approx & \one + \left(\alpha\,{\tilde j}_{\mu\nu\lambda}+ \beta\, {\tilde J}_{\mu\nu\lambda}\right) dx^\mu dx^\nu dx^\lambda.
\end{eqnarray*}
Finally, equating the coefficients of $\alpha$ and $\beta$ on both sides, the local differential Yang-Mills equations are obtained.


\newpage

\begin{thebibliography}{99}

\bibitem{wuyang69}
 T.~T.~Wu and C.~-N.~Yang,
  ``Some solutions of the classical isotopic gauge field equations,''
  In *Yang, C.N.: Selected Papers 1945-1980*, 400-405; also in *H. Mark and S. Fernbach, Properties Of Matter Under Unusual Conditions*, New York 1969, 349-345.  
 
\bibitem{wuyang75} 
  T.~T.~Wu and C.~N.~Yang,
  ``Concept of Nonintegrable Phase Factors and Global Formulation of Gauge Fields,''
  Phys.\ Rev.\ D {\bf 12}, 3845 (1975).
  doi:10.1103/PhysRevD.12.3845
  
  
  \bibitem{pappas} 
  G.~Lanyi and R.~Pappas,
  ``Nonuniqueness of the Source for Singular Gauge Fields,''
  Phys.\ Lett.\ B {\bf 68}, 436 (1977).
  doi:10.1016/0370-2693(77)90463-4
 
  
  \bibitem{oh} 
  C.~H.~Oh,
  ``Singularities And Sources Of Static Gauge Fields,''
  Phys.\ Rev.\ D {\bf 19}, 1248 (1979).
  doi:10.1103/PhysRevD.19.1248
  
  
  \bibitem{butera} 
  P.~Butera, G.~M.~Cicuta and M.~Enriotti,
  ``Singular Classical Solutions Of Euclidean Field Theories,''
  Phys.\ Rev.\ D {\bf 20}, 456 (1979).
  doi:10.1103/PhysRevD.20.456
  

\bibitem{ym1} 
  L.~A.~Ferreira and G.~Luchini, 
  ``Integral form of Yang-Mills equations and its gauge invariant conserved charges,'' 
  Phys.\ Rev.\ D {\bf 86}, 085039 (2012); 
  [arXiv:1205.2088 [hep-th]].
  
\bibitem{ym2} 
  L.~A.~Ferreira and G.~Luchini, 
  ``Gauge and Integrable Theories in Loop Spaces,''
  Nucl.\ Phys.\ B {\bf 858}, 336 (2012); 
  [arXiv:1109.2606 [hep-th]].
  
  \bibitem{thooft} 
  G.~'t Hooft,
  ``Magnetic Monopoles in Unified Gauge Theories,''
  Nucl.\ Phys.\ B {\bf 79}, 276 (1974). 
  doi:10.1016/0550-3213(74)90486-6
  
  \bibitem{polyakov} 
  A.~M.~Polyakov,
  ``Particle Spectrum in the Quantum Field Theory,''
  JETP Lett.\  {\bf 20}, 194 (1974); 
  [Pisma Zh.\ Eksp.\ Teor.\ Fiz.\  {\bf 20}, 430 (1974)].
  
  
  \bibitem{shnir} 
  Y.~M.~Shnir,
  ``Magnetic monopoles,''
  Berlin, Germany: Springer (2005) 532 p;\\
 E.~J.~Weinberg,
  ``Classical solutions in quantum field theory : Solitons and Instantons in High Energy Physics,'' Cambridge Univ. Press (2012);\\
  N.~S.~Manton and P.~Sutcliffe,
  ``Topological solitons,'' Cambridge Univ. Press (2004)
  
   \bibitem{distribution}
 Laurent Schwartz, ``Mathematics for the Physical Sciences'', Addison-Wesley (1966)\\
 I.M. Gel'fand and G.E. Shilov, ``Generalized Functions'', Academic Press (1964).

  
  \bibitem{afg1} 
  O.~Alvarez, L.~A.~Ferreira and J.~Sanchez Guillen,
  ``A New approach to integrable theories in any dimension,''
  Nucl.\ Phys.\ B {\bf 529}, 689 (1998); 
  doi:10.1016/S0550-3213(98)00400-3; 
  [hep-th/9710147].
  
  \bibitem{afg2} 
  O.~Alvarez, L.~A.~Ferreira and J.~Sanchez-Guillen,
  ``Integrable theories and loop spaces: Fundamentals, applications and new developments,''
  Int.\ J.\ Mod.\ Phys.\ A {\bf 24}, 1825 (2009); 
  doi:10.1142/S0217751X09043419; 
  [arXiv:0901.1654 [hep-th]].
  
  \bibitem{stokes} 
  I.~Arefeva, 
  ``NonAbelian Stokes formula,''
  Theor. Math. Phys. {\bf 43}, 353 (1980);
  [Teor. Mat. Fiz.  {\bf 43}, 111 (1980)].
  doi:10.1007/BF01018469\\
N.~E.~Bralic, 
  ``Exact Computation of Loop Averages in Two-Dimensional Yang-Mills Theory,''
  Phys. Rev. D {\bf 22}, 3090 (1980);
  doi:10.1103/PhysRevD.22.3090
  
   \bibitem{directtest} 
  C.~P.~Constantinidis, L.~A.~Ferreira and G.~Luchini,
  ``Direct test of the integral Yang-Mills equations through SU(2) monopoles,''
  Phys.\ Rev.\ D {\bf 96}, no. 10, 105024 (2017); 
  doi:10.1103/PhysRevD.96.105024; 
  [arXiv:1710.03359 [hep-th]].

  
 \bibitem{remark} 
  C.~P.~Constantinidis, L.~A.~Ferreira and G.~Luchini,
  ``A remark on the asymptotic form of BPS multi-dyon solutions and their conserved charges,''
  {\em Jounal of High Energy Physics} JHEP {\bf 1512}, 137 (2015); 
  doi:10.1007/JHEP12(2015)137; 
  [arXiv:1508.03049 [hep-th]].
  
   
 
  
\end{thebibliography}
\end{document}